\documentclass[twocolumn,pra,showpacs,floatfix]{revtex4}

\usepackage{amssymb}
\usepackage{amsmath}
\usepackage{amsfonts}
\usepackage{graphicx}

\begin{document}

\author{D.\ A.\ Lidar}
\affiliation{Chemical Physics Theory Group, Chemistry Department, University of Toronto,
80 Saint George Street, Toronto, Ontario M5S 3H6, Canada }
\author{J.\ H.\ Thywissen}
\affiliation{McLennan Physical Laboratories, Physics Department, University of Toronto,
60 Saint George Street, Toronto, Ontario M5S 1A7, Canada}
\title{Exponentially Localized Magnetic Fields for Single-Spin Quantum Logic
Gates}

\begin{abstract}
An infinite array of parallel current-carrying wires is known, from the
field of neutral particle optics, to produce an exponentially localized
magnetic field when the current direction is antiparallel in adjacent wires.
We show that a finite array of several tens of superconducting Nb nanowires
can produce a peak magnetic field of 10\thinspace mT that decays by a factor
of $10^{4}$ over a length scale of $500\,$nm. Such an array is readily
manufacturable with current technology, and is compatible with both
semiconductor and superconducting quantum computer architectures. A series
of such arrays can be used to individually address single single-spin or
flux qubits spaced as little as $100$\thinspace nm apart, and can lead to
quantum logic gate times of $5$\thinspace ns.
\end{abstract}

\maketitle

\section{Introduction}

Among the many and varied proposals for constructing quantum computers,
spintronic solid state devices occupy a special place because of the
prospects of integration with the existing semiconductor technological
infrastructure \cite{Awschalom:book}. At the same time, superconducting
devices have taken an early lead in demonstrating the viability of the
building blocks of quantum computing (QC) in the solid state, with recent
reports of controlled single-qubit operations and entanglement generation 
\cite{Vion:02-Yu:02-Berkley:03}.

In both the spintronics QC proposals, such as quantum dots \cite%
{Loss:98,Burkard:99,Levy:01a-Hu:01a,Friesen:02}, donor spins in Si \cite%
{Kane:98-Vrijen:00-Mozyrsky:01}, electrons on helium \cite{Lyon:03}, and the
superconducting QC proposals \cite{vdWal:00-Friedman:00-Blais:00-Makhlin:01}%
, the ability to apply highly localized and inhomogeneous magnetic fields
would be a definite advantage, if it could be done without excessive
technical difficulties. In fact the early proposals suggested manipulating
individual spin qubits using such localized magnetic fields, e.g., by a
scanning-probe tip or by coupling to an auxiliary ferromagnetic dot \cite%
{Loss:98}, but there are significant speed, controllability and other
difficulties associated with such methods. Because of these difficulties, in
particular the spintronics requirement to resolve single spins, many
alternatives to the use of localized magnetic fields have been proposed in
spin-based QC. These alternatives typically avoid the use of magnetic fields
altogether: e.g., $g$-factor engineering combined with all electrical
control \cite{Kato:03}, optical spin manipulation \cite{Piermarocchi:02}, or
encoding into the states of several spins \cite%
{Bacon:99a-Kempe:00-DiVincenzo:00a,LidarWu:01}. Other alternatives include
gate teleportation, which requires controllable exchange interactions and
certain two-spin measurements \cite{WuLidar:02a}, and qubits encoded into
antiferromagnetic spin clusters, in which case the magnetic field needs to
be controlled only over the length scale of the cluster diameter \cite%
{Meier:02}. In the context of superconducting qubits it is also possible to
avoid using localized magnetic fields by introducing an appropriate encoding 
\cite{LidarWuBlais:02}. 
\begin{figure}[b!]
%\vspace{6cm}
\par
\begin{center}
\includegraphics[height=5cm,angle=0]{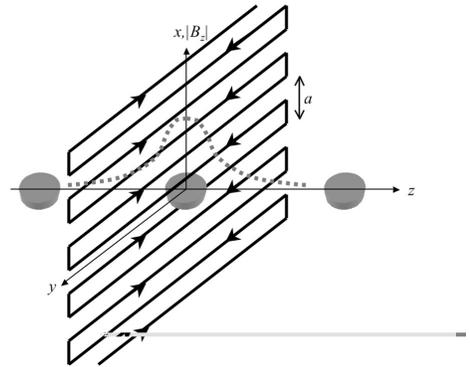}
\end{center}
\caption{Schematic depiction of quantum dots (disks) with array of current
carrying wires. Array period is $a$, current is $I$ and alternates direction
as indicated by arrows. The resulting magnetic field profile is drawn
schematically.}
\label{fig:array}
\end{figure}

Here we revisit the possibility of applying highly localized magnetic
fields. We show that a scheme inspired by magnetic mirrors for cold neutrons 
\cite{Vladimirskii:61}, and more recently cold atoms \cite%
{Opat:92,Sidorov:96,Johnson:98,Lau:99,Zabow:99} is capable of generating a
magnetic field that decays exponentially fast over a length scale comparable
to the spacing between nanofabricated quantum dots, and has strength and
switching times that are compatible with QC given available estimates of
decoherence times. Our scheme uses arrays of parallel current-carrying
wires, that is readily implementable with currently available nanotechnology 
\cite{Melosh:03,Awschalom:book}, and appears well suited for integration
with quantum dot nanofabrication methods, as well as with superconducting
flux qubits and spin-cluster qubits, where the length scales are larger.
Thus we believe that QC with localized magnetic fields deserves a fresh
look. 

\section{Exponentially localized magnetic field from an infinite wire array}

In order to have a concrete application in mind we shall from now on refer
to semiconductor quantum dot spin-qubits \cite%
{Loss:98,Burkard:99,Levy:01a-Hu:01a,Friesen:02}. However, our results are
equally applicable to other qubits that are manipulated by localized
magnetic fields, such as superconducting flux qubits \cite%
{vdWal:00-Friedman:00-Blais:00-Makhlin:01}. The first requirement for
single-spin magnetic resolution is a magnetic field profile that decays
exponentially fast over length scales comparable with the inter-spin
spacing. We will now show, in close analogy to results from magnetic
mirrors, how to produce such an exponentially localized magnetic field. The
basic design is one of an array of parallel current carrying wires, with the
current direction alternating from wire to wire: see Fig.~\ref{fig:array}.

We first consider the idealized case of infinitely long wires. In this case
the field can be calculated analytically (see also \cite%
{Vladimirskii:61,Opat:92}). Let $B_{\alpha }^{N}$ be the magnetic field in
the $\alpha =x,z$ direction generated by $N$ infinitely long wires. We add
magnetic field contributions from each wire, to get the field components from $N$ wire pairs: 
\begin{widetext}
%\vspace{-1cm}
\begin{eqnarray}
B_{z}(x,z) &=&\frac{\mu _{0}I}{2\pi a}\sum_{n=0}^{N-1}{\left( -1\right) }%
^{n}\,\left( \frac{(\frac{1}{2}+n)/2-x}{{\left( (\frac{1}{2}+n)/2-x\right) }%
^{2}+z^{2}} +\frac{(\frac{1}{2}+n)/2+x}{{\left( (\frac{1}{2}+n)/2+x\right) }%
^{2}+z^{2}}\right) , \\
B_{x}(x,z) &=&\frac{\mu _{0}I}{2\pi a}\sum_{n=0}^{N-1}{\left( -1\right) }%
^{n}\,\left( \frac{z}{{\left( (\frac{1}{2}+n)/2-x\right) }^{2}+z^{2}} -\frac{z%
}{{\left( (\frac{1}{2}+n)/2+x\right) }^{2}+z^{2}}\right) ,
\end{eqnarray}
\end{widetext}
where $I$ is the current through each wire and $k=2\pi /a$ is the reciprocal
array constant. This sum can be computed analytically in the limit $%
N\rightarrow \infty $ using the residue theorem result  

\begin{figure}[b!]
%\vspace{7cm}  
\includegraphics[height=4.5cm,angle=0]{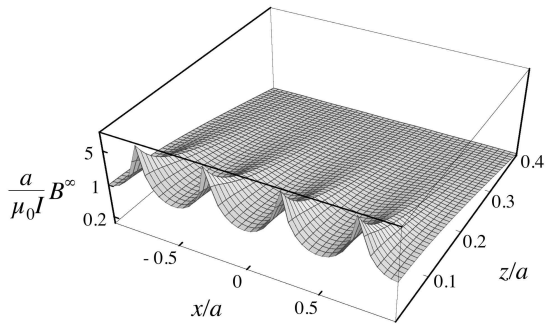}
\caption{The magnetic field magnitude from a infinite array of infinitely
long wires (\textquotedblleft doubly infinite\textquotedblright ), on a
logarithmic scale. Note that the field is flat between the wires, and decays
exponentially in $z$.}
\label{fig:Bz}
\end{figure}
\begin{equation*}
\sum_{n=-\infty }^{\infty }(-1)^{n}f(n)=-\pi \sum_{\zeta _{k}}\frac{1}{\sin
\pi \zeta _{k}}\mathrm{Res}(f,\zeta _{k}),
\end{equation*}
where $\zeta _{k}$ are the poles of $f(\zeta )$, yielding 
\begin{eqnarray}
B_{x}^{\infty }(x,y,z) &=&\frac{\mu _{0}I}{a}\frac{2\,\sin (k\,x)\,\sinh
(k\,z)}{\cos (2\,k\,x)+\cosh (2\,k\,z)},  \label{eq:Bx} \\
B_{z}^{\infty }(x,y,z) &=&\frac{\mu _{0}I}{a}\frac{2\,\cos (k\,x)\,\cosh
(k\,z)}{\cos (2\,k\,x)+\cosh (2\,k\,z)}  \notag \\
&\overset{x=0}{=}&\frac{\mu _{0}I}{a}\mathrm{sech}(kz)\overset{z\gg a}{%
\rightarrow }\frac{2\mu _{0}I}{a}{e}^{-k|z|}.  \label{eq:Bz}
\end{eqnarray}%
The $z$-component result shows the basic point: an exponentially localized
magnetic field can be generated using a wire array. The flat top of the $%
\mathrm{sech}$ profile is a useful design feature, since it implies no
exponential sensitivity in the range $z\lesssim a$ \cite{Friesen:02}. The
field magnitude 
\begin{equation}
B^{\infty }=\left\vert \mathbf{B^{\infty }}\right\vert =\sqrt{2}\left[ {\cos
(2kx)+\cosh (2kz)}\right] ^{-1/2}
\end{equation}
oscillates with period $a$ in the $x$-direction: see Fig.~\ref{fig:Bz}.

\section{Multiple Arrays}

For the purposes of QC, we should ideally be able to address each spin
separately. To this end we propose to center a separate wire array on each
quantum dot. Then, as long as the dots are spaced on the order of the
lattice constant $a$, we have exponentially sensitive addressability of each
dot. The introduction of multiple arrays is useful in another respect: we
can adjust the magnitudes and directions of currents in different arrays so
as to exactly cancel the field at all (or only some) other dots except the
desired one (or ones). To see this, let $b(z)\equiv B_{z}^{\infty }(z)/I=%
\frac{\mu _{0}}{a}\mathrm{sech}{kz}$. The field at position $z$ from $K$
arrays of wires, with the $j$th array having current $I_{j}$ and
intersecting the $z$-axis at position $z_{j}$ (typically the center of one
of the dots), is: 
\begin{equation}
B^{\{K\}}(z)=\sum_{j=-(K-1)/2}^{(K-1)/2}I_{j}b(z-z_{j}).
\end{equation}%
Suppose we wish the field to have magnitude $c_{j}$ at position $z_{j}$.
Formally, we need to solve: 
\begin{equation}
B^{\{K\}}(z_{j})=c_{j},\quad j\in \lbrack -(K-1)/2,(K-1)/2].
\end{equation}%
This is a linear system of $K$ equations in the $K$ unknowns $I_{j}$, so it
can always be solved in terms of the $K$ positions $z_{j}$. E.g., the fields
with and without the correction are shown, for $K=5$, in Fig.~\ref%
{fig:cancelB}. 
\begin{figure}[t!]
%\vspace{7cm}  
\includegraphics[height=3.7cm,angle=0]{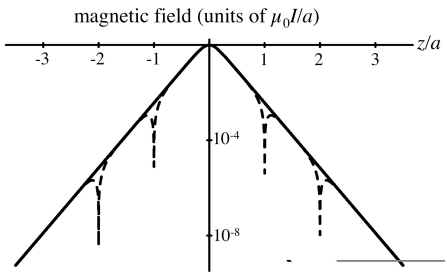}
\caption{Magnetic field generated by a single doubly infinite wire array
centered at $z=0$ (solid), \textit{vs} the field generated by $K=5$ such
arrays (dashed), with currents chosen to cancel the field at positions $z=\pm
2a,\pm a$.}
\label{fig:cancelB}
\end{figure}

\section{Finite size effects}

For a finite system ($N<\infty $ wires, finite length and thickness,
non-ideal shape, etc.), we can only expect the above results to hold to an
approximation. Much of the theory of corrections to finite size effects has
already been worked out in \cite{Sidorov:96,Zabow:99}, in the context of
atomic mirrors. Since for atomic mirrors the primary concern is specular
reflection, there the focus was on reducing the variation of the field
magnitude in the planes parallel to the wire arrays. For us this criterion
is unimportant; instead, our focus is on making the field as localized as
possible along the $z$-axis.

The first important conclusion in the case of a finite number of wires $N$
is that there is a transition from exponential to quadratic (i. e., $1/z^2$)
decay. \cite{Zabow:99} The transition takes place at the inflection point
(zero second derivative) of $B^{N}(z)$; however, this is difficult to obtain
analytically. To roughly estimate the transition point we compute where the
field from a single pair of wires, positioned at the edge of an array of $N$
wires ($y=\pm Na/4$), generates a field of magnitude equal to that from an
infinite array: 
\begin{equation}
\frac{\mu _{0}I}{\pi }\frac{Na/4}{z^{2}+(Na/4)^{2}}=\frac{\mu _{0}I}{a}{e}
^{-2\pi z/a}.
\end{equation}
Since the transition happens for $x/a\ll N$ we neglect $x/a$ in the
denominator. The solution is then 
\begin{equation}
z_{t} \approx \frac{\pi a}{20} \ln (\frac{\pi N}{4})  \label{eq:zt}
\end{equation}
Numerical calculations show that Eq.~(\ref{eq:zt}) overestimates the
position of the transition point by about a factor of $5$; however, after
this correction is made, analytics and numerics agree well across several
orders of magnitude of $N$.

This logarithmic dependence might appear to pose a severe scalability
constraint on our method. However, this is not the case when we take into
account the threshold for fault tolerant quantum error correction \cite%
{Steane:02}. For, it follows from the threshold result that we only need to
make the ratio of the peak field (applied to the desired spin) to the
residual field smaller than, say $10^{-4}$. The crucial question thus
becomes for what $N$ this can be achieved, and this brings us to the idea of
\textquotedblleft endcaps\textquotedblright . 

As observed in \cite{Sidorov:96,Lau:99,Zabow:99}, near the center of the
array the magnetic field that would be produced by the semi-infinite array
of ``missing'' wires is the same (to first order in $4/N$) as that of a pair
of wires carrying current $I_{\mathrm{cap}}=I/2$ and placed with their
centers shifted by $a/4$ from the outermost wires in the array. Thus, to
cancel this field, one can simply place two ``endcap'' wires carrying
currents $\pm I/2$ at these positions. In the context of atomic mirrors this
is important to improve flatness, and hence specular reflection. In \cite%
{Zabow:99} it was observed that flatness can be further enhanced by using an 
\emph{odd} number of wires.

Some of these schemes can be used to improve \emph{field localization}, a
criterion not considered orginally. For instance, we find that the number of
wires can be reduced drastically -- from $N\approx 10^{4}$ to as few as $N=22
$ wires -- when endcaps are used to achieve a residual field smaller than $%
10^{-4}$. By contrast, using an odd $N$ is disastrous for localization: for
example, residual fields appear at the $3\%$ level for $N=23$, even
including endcap correction. Higher (even) wire number increases the
robustness of the cancellation against experimental uncertainty in the
current and position of the endcaps: $N=30$ is required to maintain $B/\mu
_{0}I\leq 10^{-4}$ for a fractional current variation of $\pm 10^{-3}$ and a
positional uncertainty of $\pm 2.5$\thinspace nm, as shown in Fig. \ref%
{fig:optimized}. Finally, we note that the corrections arising from the
finite length of the wires and the short, perpendicular connecting wires,
can also be compensated for by the use of judiciously placed compensating
wires \cite{Zabow:99}.

\section{Feasibility and implementation considerations}

\begin{figure}[t!]
%\vspace{6cm} 
\includegraphics[height=4.5cm,angle=0]{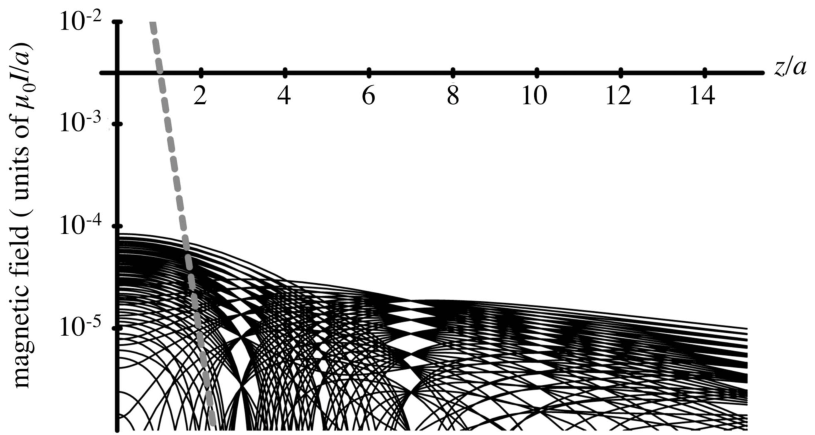}
\caption{Deviation $\left\vert {\mathbf{B}^{\infty }}-{\mathbf{B}^{30}}%
\right\vert $ of the field of a finite array (solid lines) from the field $%
\left\vert {\mathbf{B}^{\infty }}\right\vert $ of an infinite array (grey dashed line,
shown for comparison). This array has $N=30$ wires (including two endcap
wires), such that main wires are at $\pm a/4,\pm 3a/4,\ldots \pm 27a/4$, and
endcap wires are at $\pm 7a$, carrying current $I_{\mathrm{cap}}=I/2$. Field
deviations are plotted for a range of endcap currents ($\pm 10^{-3}I_{%
\mathrm{cap}}$) and positions ($\pm a/100=2.5$\thinspace nm). Note that the
residual field never exceeds the threshold of $10^{-4}$.}
\label{fig:optimized}
\end{figure}
We now come to estimates of whether the fields and size scales required are
feasible in practice. Let us first calculate the magnetic field strength
required for single-qubit operations. A spin can be rotated by a relative
angle $\phi =g\mu _{B}B\tau /2\hbar $ by turning on the field $B$ for a time 
$\tau $ (where $g\approx 1$ is the $g$-factor, and $\mu _{B}$ is the Bohr
magneton). Recent estimates of dephasing times are $50\,\mu \mathrm{s}$ for
electron spins in GaAs quantum dots (a calculation, assuming spectral
diffusion is dominant) \cite{deSousa:03}, and a measurement of $60$%
\thinspace ms for $T_{2}$ of phosphorus donors in Si \cite{Tyryshkin:03}. If
we use the more pessimistic of these numbers, and assume a fault tolerance
threshold of $10^{-4}$ for QC \cite{Steane:02}, we may estimate the desired
operation time as $\tau \sim 10^{-4}\times 50\,\mu $s $=5$\thinspace ns for
an angle $\phi =\pi /2$. Thus the desired field strength is $B=2\hbar \phi
/g\mu _{B}\tau \approx 7$\thinspace mT.

To evaluate the feasibility of such a specification, we consider an array
with periodicity $a=250$\thinspace nm, wire radius $r=50$\thinspace nm, $N=32
$ wires, and length $L_{y}=10\,\mu $m along the $y$-direction. These
dimensions are compatible with the $100$\thinspace nm length scales of
quantum dots \cite{Friesen:02}. In order to reach the desired field strength
of $7$\thinspace mT, $I=aB_{z}^{\infty }/\mu _{0}\approx 1.4$\thinspace mA
would be required. However, decoherence due to heating with such a current
could be a major issue. An upper bound estimate for the required temperature 
$T$ can be given by $T\ll E_{Z}/k_{B}$, where $E_{Z}=g\mu _{B}B/2$ is the
Zeeman splitting of the spins in the applied magnetic field $B$. In our
case, we have constrained $B$ by the gate time $\tau $, so we can write $%
k_{B}T\ll \pi \hbar /2\tau $, or $T\ll 2.4$\thinspace mK for $\tau =5$%
\thinspace ns. This is feasible with dilution refrigeration technology if
the heat load is on the order of $100$\thinspace pW \cite{comment},
comparable to the dissipation of quantum dots \cite{Friesen:02}.

For normal metal wires, such a heat load restriction is prohibitive. The
power $P$ dissipated is $P=j^{2}\rho A\ell ,$ where $j$ is the current
density, $\ell =N(2L_{y}+a)$ is the total wire length, and $\rho $ is the
resistivity. Below $10$\thinspace K, oxygen-free copper can have $\rho
\approx 3\times 10^{-11}$\thinspace $\Omega $\thinspace m. \cite{Fickett:83}
At $P\leq 100$\thinspace pW, $I\leq 3.6\,\mu $A, which would give $\tau \geq
2.0\,\mu $s -- nearly a thousand times slower than our original goal.
Although this would be acceptably fast if the decoherence time were $60$%
\thinspace ms, as in \cite{Tyryshkin:03}, the high-purity $\rho $ we have
used is optimistic for nanofabricated wires, and the heat generated would
increase linearly with the number of qubits manipulated.

One can circumvent resistive heating by using superconductors. A wire with
radius $r \lesssim \xi_0$, where $\xi_0$ is the coherence length, can also
avoid heating mechanisms associated with vortex movement through the
superconductor. For $r \sim \lambda$ or smaller, where $\lambda$ is the
penetration depth, the critical current density is due to depairing: $j_c =
(2/3)^{3/2} B_c / (\mu_0 \lambda)$ \cite{Tinkham:book}. These constraints
are compatible when $\lambda \lesssim \xi_0$, i.e., mostly type I
superconductors. For Nb, $B_{c1}=0.206$\,T, $\lambda=52$\,nm, and $\xi_0 = 39
$\,nm \cite{Poole}, so $j_c = 1.7\times10^{12}$\,A/m$^2$. Note that
high-temperature superconductors typically have higher $\lambda$ and thus
lower $j_c$. In any case, the critical current density of Nb is more than is
required: a gate time of $\tau=5$\,ns would require $j \approx 0.1 j_c$.
\section{Conclusions and Extrapolations}

In conclusion, our results indicate that QC with localized magnetic fields
deserves renewed consideration. We have shown that a method to produce
exponentially decaying magnetic fields using an array of current-carrying
wires, known in the cold neutron and atom optics communities, is adaptable
to solid state quantum computer implementations. Our estimates indicate that
in all respects the method is technologically feasible, provided
superconducting wires with sufficiently high critical current density (such
as Nb) are used.

Our work is motivated by the quest to perform single-spin or flux-qubit
rotations, which is a component of a universal set of quantum logic gates.
The geometry shown in Fig.~\ref{fig:array} yields a field that is localized
in the $z$-direction; in order to perform arbitrary single-qubit rotations
we need to localize the field along another, perpendicular direction. An
independent field vector can be produced by a second set of interleaved
arrays placed at 45$^{\circ }$ with respect to the original arrays, with
current flowing along $(\mathbf{z}+\mathbf{y})/\sqrt{2}$. With a judicious
array placement and $na=d/\sqrt{2}$, for qubit spacing $d$ and any integer $n
$, the field direction will be along $\mathbf{x}$ at all qubits (this design
will need to be optimized similarly to our considerations above -- an issue
we do not intend to address here). The additional spatial constraints would
require only a four-fold increase in $a$ for the same currents and wire
sizes. If on the other hand, introducing a second array is undesirable,
\textquotedblleft software\textquotedblright\ solutions using recoupling and
encoding techniques have been developed to still allow for universal QC \cite%
{LidarWu:01}. These techniques would be considerably simplified by the
ability to perform single-qubit operations along one direction.

\textit{Acknowledgments}. We thank Dr. Alexandre Zagoskin, Dr. Brian Statt,
and Dr. John Wei for useful discussions. D.A.L.'s research is sponsored by
the Defense Advanced Research Projects Agency under the QuIST program and
managed by the Air Force Research Laboratory (AFOSR), under agreement
F49620-01-1-0468. D.A.L. further gratefully acknowledges the Alfred P. Sloan
Foundation for a Research Fellowship. J.H.T. thanks the Canada Research
Chairs program for financial support.

\end{document}